\documentclass[superscriptaddress,twocolumn,showpacs,amsmath,amssymb]{revtex4}

\usepackage[dvips]{graphicx}
\usepackage{graphicx}

\renewcommand{\vec}[1]{\mbox{\boldmath $#1$}}

\begin{document}

\title{
Role of non-collective excitations in heavy-ion 
fusion reactions and quasi-elastic
scattering around the Coulomb barrier}

\author{S. Yusa}
\affiliation{
Department of Physics, Tohoku University, Sendai 980-8578,  Japan} 

\author{K. Hagino}
\affiliation{
Department of Physics, Tohoku University, Sendai 980-8578,  Japan} 

\author{N. Rowley}
\affiliation{
Institut de Physique Nucl\'{e}aire, UMR 8608, CNRS-IN2P3 et Universit\'{e}
de Paris Sud, 91406 Orsay Cedex, France}

\begin{abstract}
Despite the supposed simplicity of double-closed shell nuclei, conventional
coupled-channels calculations, that include all of the known collective states of the target
and projectile, give a poor fit to the fusion cross section for the $^{16}$O + $^{208}$Pb system. 
The discrepancies are highlighted through the experimental barrier distribution and 
logarithmic derivative, that are both well defined by the precise experimental fusion data available.
In order to broaden our search for possible causes for this anomaly, we revisit this system
and include in our calculations a large number of non-collective states of the target,
whose spin, parity, excitation energy and deformation paramter are known from high-precision proton inelastic-scattering
measurements. Although the new coupled-channels calculations modify the barrier distribution,
the disagreemnt with experiment remains both for fusion and for quasi-elastic (QE) scattering.
We find that the Q-value distributions for large-angle QE scattering become rapidly more important
as the incident energy increases, reflecting the trend of the experimental data. The mass-number
dependence of the non-collective excitations is discussed.
\end{abstract}


\maketitle

\section{Introduction}

In heavy-ion reactions at energies close to the Coulomb barrier, 
couplings between the relative
motion and the internal degrees of freedom in the colliding nuclei
play a crucial role. 
A well known example is the enhancement of 
subbarrier fusion cross sections, 
compared to the predictions of a simple potential model, 
due to strong couplings to collective excitations
~\cite{dasgupta,BT98}. 
Coupled-channels analyses have been performed for various 
heavy-ion systems taking into account such coupling effects and
have successfully accounted for experimental data 
for fusion reactions as well as quasi-elastic scattering at backward angles~\cite{dasgupta}.

Conventionally, a few low-lying collective excitations, such as vibrational
modes in spherical nuclei or rotational excitations in deformed nuclei, 
as well as few-nucleon transfer channels  
have been taken into account
in the coupled-channels calculations. 
In the eigen-channel representation, channel-coupling effects lead to 
a distribution of potential barriers~\cite{DLW83}, and it
has been well established that the barrier distribution can be directly extracted from
experimental fusion and quasi-elastic cross sections.
For fusion reactions, the barrier distribution is
defined as the second derivative 
of the product of center-of-mass energy $E_{\rm cm}$ and fusion cross section $\sigma_{\rm fus}$ with respect 
to $E_{\rm cm}$, that is,
$d^2(E_{\rm cm}\sigma_{\rm fus})/dE_{\rm cm}^2$~\cite{RSS91,L95}.
For quasi-elastic scattering, it is defined as the first derivative of the
ratio of the backward quasi-elastic scattering cross section to 
the Rutherford cross section with respect
to center-of-mass energy, that is $-d\left(\sigma_{\rm
qel}(\theta=\pi)/\sigma_{\rm R}(\theta = \pi)\right)/dE_{\rm
cm}$~\cite{timmers,HR04}.
The fusion and quasi-elastic barrier distributions have been found to behave
in a similar way to each other, though the quasi-elastic barrier distribution tends to be more
smeared~\cite{timmers,zamrun}.
These quantities are known to be sensitive to the channel coupling effects~\cite{dasgupta,BT98,L95}.
They can also serve for the determination of deformation parameters
~\cite{LRL93}.

Although the coupled-channels method appears successful for heavy-ion 
subbarrier fusion reactions, there is a long standing problem of the method
that, in order to reproduce experimental fusion data,
a significantly larger value of the  surface diffuseness of the nuclear potential
is required, compared to the value found from fitting the 
scattering process~\cite{MHD07, N04}. 
Furthermore, some recently obtained experimental data cannot be accounted for 
by the conventional coupled-channels method. 
For example, fusion cross sections for several systems 
at deep subbarrier energies have  turned out to be 
much smaller than the predictions of conventional 
coupled-channels calculations
~\cite{J02,JRJ04,D07,S08}.  
Another example is the quasi-elastic barrier distribution for the
$^{20}$Ne + $^{90,92}$Zr systems.
The conventional coupled-channels analysis that takes
into account only the collective excitations of the colliding nuclei
fails to 
explain the difference in the experimental 
quasi-elastic barrier distributions 
of these two systems~\cite{piasecki}.
That is, 
although the experimental barrier distribution for the $^{20}$Ne + $^{92}$Zr
system is much more smeared than that for the $^{20}$Ne + $^{90}$Zr system, 
the coupled-channels calculations 
yield similar barrier distributions for both systems due to the much
larger deformation of the $^{20}$Ne nucleus. 
One possible reason for the smearing may be 
the effect of transfer
reactions~\cite{TCS97}.
However the total transfer cross sections for these systems have been found to be almost
the same~\cite{piasecki}. Therefore the difference between the barrier distributions 
has been conjectured to arise from non-collective excitations, that are not 
taken into account explicitly in the coupled-channels calculations.

In order to 
discuss how non-collective excitations affect low-energy 
heavy-ion reactions, in Ref.~\cite{YHR10} 
we have solved schematic one-dimensional coupled-channels equations 
with a gaussian potential barrier. 
There are several ways to describe the non-collective 
degrees of freedom~\cite{D10, akw1,akw2,akw3,akw4,Z90}. 
In Ref.~\cite{YHR10}, we employed 
the random matrix theory, that has been applied to deep 
inelastic collisions by Agassi {\it et al.}~\cite{akw1,akw2,akw3,akw4}. 
We have shown that, by including non-collective excitations, 
the barrier penetration
probabilities are suppressed at energies above the barrier leading to a 
smeared barrier distribution.

In this paper, we carry out a similar analysis in three dimensions, 
using realistic spectra for the non-collective excitations. 
For this purpose, we choose 
the $^{16}$O + $^{208}$Pb system. 
In addition to the experimental data 
for subbarrier and deep subbarrier fusion 
reactions~\cite{MBD99,DHD07}, 
as well as quasi-elastic scattering for this system 
at energies near the Coulomb barrier~\cite{T97,T96,evers,lin}, 
almost all of the excited states of $^{208}$Pb up to 7.5~MeV have 
been identified (spin, parity, excitation energy and deformation parameter)
from high-resolution proton 
inelastic scattering measurements~\cite{WCHM75,LBF73}. 
We will include these $^{208}$Pb excited states in our coupled-channels 
calculations and discuss the role of non-collective excitations in 
heavy-ion reactions around the Coulomb barrier. 

Note that 
a satisfactory description of the fusion cross
sections as well as the fusion barrier distribution 
has not yet been obtained for this system with 
the conventional coupled-channels 
calculations~\cite{MBD99,EM07}. That is, 
the height of the main peak in the barrier distribution is 
overestimated by the coupled-channels calculations 
(see also Ref.~\cite{IHI09}).  
Another motivation to choose 
the $^{16}$O + $^{208}$Pb system in the present study is, 
therefore, to see 
whether the non-collective excitations improve the
agreement of the coupled-channels calculation with the experimental data.

In addition to our calculation for the $^{16}$O + $^{208}$Pb system,
we also 
study 
$^{32}$S + $^{208}$Pb and $^{40}$Ca + $^{208}$Pb to
investigate the dependence of the effect of the non-collective excitations
on the mass number of the projectile.

The paper is organized as follows.
In Sec. II, we explain the coupled-channels formalism and how to
describe the couplings to the non-collective excitations.
In Sec. III, we apply the coupled-channels formalism to 
the fusion and quasi-elastic scattering of the 
$^{16}$O + $^{208}$Pb system.  
We will discuss fusion and quasi-elastic 
cross sections, barrier distributions, as well as the energy 
dependence of the Q-value distributions for quasi-elastic scattering. 
We also investigate the fusion reaction for the $^{32}$S + $^{208}$Pb and
$^{40}$Ca + $^{208}$Pb systems and 
discuss the mass-number dependence of the non-collective effects. 
In Sec. IV, we summarize the paper.

\section{Coupled-channels method}

In order to take into account excitations of the colliding nuclei during 
the fusion and scattering processes, 
we assume the following Hamiltonian:
\begin{eqnarray}
  H = -\frac{\hbar^2}{2\mu}\nabla^2 + V_{\rm rel}(r) + H_0(\xi) 
+ V_{\rm coup}(\vec{r},\xi), 
\end{eqnarray}
where $\vec{r}$ is the coordinate for the relative motion between the 
projectile and
the target nuclei, and $\mu$ is the reduced mass.  
$H_0(\xi)$ is the intrinsic Hamiltonian, 
for which 
we consider vibrational excitations of the colliding nuclei, 
$\xi$ representing the internal degrees of
freedom. 
$V_{\rm coup}(\vec{r},\xi)$ is the coupling Hamiltonian between 
the relative motion and the intrinsic degrees of freedom. 
$V_{\rm rel}(r)$ is the optical potential for the relative motion, 
that is given as a sum of the Coulomb and nuclear potentials, 
\begin{eqnarray}
  V_{\rm rel}(r) = \frac{Z_{\rm P}Z_{\rm T}e^2}{r}
           &-& \frac{V_0}{1+{\rm exp}\left[(r - R_{\rm N})/a\right]} \\ \nonumber
           &-& i \frac{W_0}{1+{\rm exp}\left[(r - R_{\rm W})/a_{\rm W}\right]}. 
\end{eqnarray}
Here, we have adopted the Woods-Saxon form for the nuclear potential.

The coupled-channels equations are obtained by expanding the total wave function
in terms of the eigenfunctions of $H_0(\xi)$. 
This leads to 
\begin{eqnarray} 
 \left[-\frac{\hbar^2}{2\mu}\frac{d^2}{dr^2} +  \frac{J(J+1)\hbar^2}{2\mu r^2}
+ V_{\rm rel}(r) + \epsilon_n - E\right]u_n^{J}(r) &&\\
 +  \sum_m V_{nm}(r)u_m^{J}(r) = 0, 
\end{eqnarray}
where, $\epsilon_n$ is the excitation energy for the $n$-th channel.
In deriving these equations, we have employed the isocentrifugal 
approximation~\cite{LR84,NRL86,NBT86,ELP87,T87,TMBR91,TAB92,AA94,GAN94,HTBB95} and replaced 
the angular momentum for the relative motion 
by the total angular momentum, $J$.
This approximation has been found to be valid for 
heavy-ion systems~\cite{T87}, and reduces considerably the 
dimensions of the coupled-channels problem. 

We impose the following boundary conditions in solving the coupled-channels
equations:
\begin{eqnarray}
  u_n^{J}(r) \rightarrow H_{J}^{(-)}(k_n r)\delta_{n,0} -
  \sqrt{\frac{k_0}{k_n}}S_n^{J}H_{J}^{(+)}(k_n r), 
\end{eqnarray}
for $r \rightarrow \infty$, together with regularity at the origin.
Here, $k_n = \sqrt{2\mu(E - \epsilon_n)/\hbar^2}$ is the wave number for the
$n$-th channel, where $n=0$ 
represents the entrance channel. $S_n^{J}$ is the nuclear
$S$-matrix, and $H_{J}^{(-)}(kr)$ and $H_{J}^{(+)}(kr)$ are the incoming and
the outgoing Coulomb wave functions, respectively. 
Using the $S$-matrix, 
the fusion cross sections are calculated as 
\begin{eqnarray}
\sigma_{\rm fus}(E) = \frac{\pi}{k_0^2}\sum_J(2J+1)
\left(1 - \left|\sum_nS_{n}^{J}\right|^2\right).
\end{eqnarray}
On the other hand, the differential cross sections 
for the channel $n$ are given by
\begin{eqnarray}
\frac{d\sigma_{n}}{d\Omega} = \frac{k_n}{k_0}|f_{n}(\theta)|, 
\end{eqnarray}
with
\begin{eqnarray}
f_n(\theta) &=& \frac{1}{2i\sqrt{k_0k_n}}\sum_{J} 
e^{i\left[\sigma_{J}(E)+\sigma_{J}(E-\epsilon_n)\right]} \\ \nonumber
&&\times (2J+1)P_{J}({\rm cos}\theta)(S_{n}^{J} - \delta_{n,0}) \\ \nonumber
&& + f_{\rm C}(\theta)\delta_{n,0}, 
\end{eqnarray}
where $\sigma_{J}(E)$ and $f_{\rm C}(\theta)$ are the Coulomb phase shift and
the Coulomb scattering amplitude, respectively.
The quasi-elastic scattering cross sections are calculated according to
\begin{eqnarray}
\sigma_{\rm qel}(E,\theta) = \sum_n \frac{d\sigma_n}{d\Omega}(E,\theta). 
\end{eqnarray}

\section{Results}

Let us now numerically solve the coupled-channels equations for the 
$^{16}$O+$^{208}$Pb system. 
For the coupling to the collective excitations, we take into account 
the vibrational 3$^-$ state at 2.615~MeV, 5$^-$ state at 3.198~MeV, 
and 2$^+$ state at 4.085~MeV
in $^{208}$Pb as well as the 3$^-$ state at 6.13~MeV in $^{16}$O.
The deformation parameters are estimated from the measured 
electromagnetic transition probabilities, that is, 
$\beta_3 (^{208}$Pb) = 0.122, 
$\beta_5 (^{208}$Pb) = 0.058, 
$\beta_2 (^{208}$Pb) = 0.058, 
$\beta_3 (^{16}$O) = 0.733, 
together with a radius parameter 
of $r_0$=1.2~fm. 
In addition to these collective vibrational states, we also include
70 non-collective states in $^{208}$Pb below 7.382~MeV,  
whose excitation energies, multipolarities, and
deformation parameters are taken from the high-resolution proton
inelastic scattering measurements in Ref.~\cite{WCHM75}.
We take into account the mutual excitations of the $^{208}$Pb and the 
$^{16}$O nuclei. 

For the nuclear potential, 
we use the same geometry as that in 
Ref.~\cite{evers}, 
where the parameters were obtained by fitting the coupled-channels 
calculations 
to the experimental quasi-elastic scattering cross sections.
This potential has a surface diffuseness parameter of $a=0.671$~fm. 
Since our calculation takes into account the 3$^{-}$ state in
$^{16}$O, that was not included in Ref.~\cite{evers}, 
we modify the potential depth from 853~MeV to 550~MeV in order to 
compensate the adiabatic potential 
renormalization~\cite{THAB94}. 
For the form factors of the non-collective couplings, 
for simplicity we take the same geometry as that for the collective 
couplings. 
For the non-collective excitations,
we include only the couplings between the ground state and the 
non-collective states, and neglect 
the couplings among the non-collective excitations and the couplings
between the collective and the non-collective states.

\subsection{Single phonon calculations}
We first show the results for the calculation that does not take into account
double octupole phonon states in the $^{208}$Pb. In this case, the number of
channels amounts to 146 in the isocentrifugal approximation. 

\begin{figure}[t]
    \includegraphics[clip,width=71.46mm,height=156.20mm]{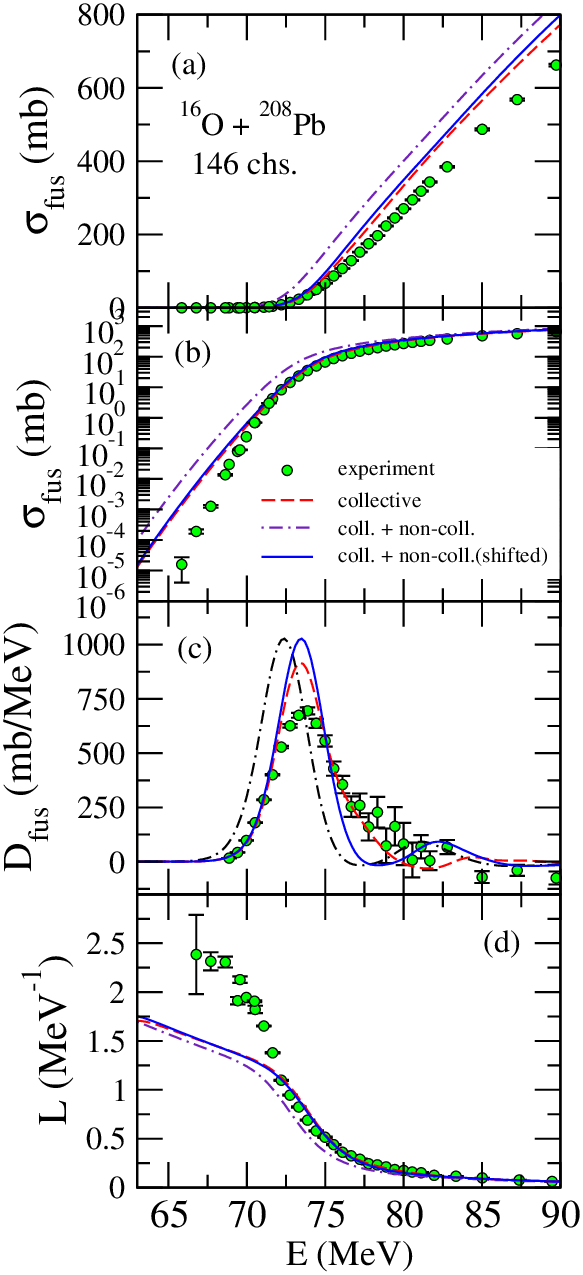}
    \caption{(Color online) 
The fusion cross sections (Fig.~1(a) and 1(b)),  
the fusion barrier distribution, $D_{\rm fus} = d^2(E\sigma_{\rm fus})/dE^2$, 
(Fig.~1(c)), and the logarithmic slope, 
$L(E) = d[{\rm ln}(E\sigma_{\rm fus})]/dE$, (Fig.~1(d)), 
for the $^{16}$O + $^{208}$Pb reaction. 
The fusion cross sections are plotted both on linear and logarithmic 
scales in Figs.~1(a) and 1(b), respectively. 
The dashed lines are obtained by taking into account only the 
collective excitations of $^{16}$O and $^{208}$Pb, 
while the dot-dashed lines take into account the 
non-collective excitations of $^{208}$Pb in addition to the collective
excitations. The solid lines are the same as the 
dot-dashed lines, but shifted in energy. 
The experimental data are taken from Refs.~\cite{MBD99,DHD07}.}
    \label{fig:fusion}
\end{figure}

Figures \ref{fig:fusion}(a) and 
\ref{fig:fusion}(b) 
show the fusion cross sections 
thus obtained. They are 
plotted both on a linear scale (Fig.~\ref{fig:fusion}(a)) 
and on a logarithmic scale (Fig.~\ref{fig:fusion}(b)).
The corresponding barrier distributions, 
$D_{\rm fus} = d^2(E\sigma_{\rm fus})/dE^2$, are plotted in 
Fig.~\ref{fig:fusion}(c).
The experimental data are taken from Refs.~\cite{MBD99,DHD07}.
The dashed lines are obtained by taking into account only
the collective excitations of $^{208}$Pb and $^{16}$O, 
while the dot-dashed
lines take into account also the non-collective excitations of $^{208}$Pb. 
One immediately sees that 
the main peak in the barrier distribution is shifted in energy 
due to the non-collective excitations 
towards low energy and consequently the fusion 
cross sections are enhanced. 
This can be understood in terms of the adiabatic potential
renormalization because the excitation energies for the 
non-collective excitations
are relatively large. 
One can also see that the non-collective excitations do not 
alter much the energy dependence of the fusion cross sections, as 
can be seen more clearly by shifting 
the dot-dashed lines in energy as shown in Fig.~\ref{fig:fusion} by the 
solid lines. 
As a consequence, the non-collective excitations 
hardly modify the behavior of 
the logarithmic slope, $L(E) = d[{\rm ln}(E\sigma_{\rm fus})]/dE$ (see 
Fig.~\ref{fig:fusion}(d)). 
That is, 
the calculations with only the collective excitations do not 
account for the observed large logarithmic slope at deep 
subbarrier energies. This 
remains the same even if the non-collective excitations are 
taken into account. 
This indicates that 
the deep sub-barrier hindrance of fusion cross sections cannot be 
explained simply with the non-collective excitations 
in each of the colliding nuclei; 
some other mechanism, such as non-collective excitations 
of the one-body system after the touching of the colliding nuclei, 
has to be considered~\cite{IHI09}. 

As mentioned in Sec.~I, 
it is known that the calculation with only collective excitations
does not reproduce well the experimental barrier distribution for 
this system~\cite{MBD99}. 
That is, the coupled-channels calculation yields 
too high a main peak in the barrier distribution. 
We find that the non-collective excitations are not helpful in this 
respect, as shown in Fig.~\ref{fig:fusion}(c). 
The non-collective excitations rather 
smear the barrier distribution 
at energies around 78~MeV~\cite{YHR10}, 
and the agreement is somewhat worsened. 
Clearly, one needs other mechanisms in order to 
reproduce the experimental barrier distribution for this system. 
In this connection, in the next subsection, we will investigate the effect of 
double octupole phonon excitations in $^{208}$Pb. 

\begin{figure}[t]
    \includegraphics[clip,width=85mm,height=103.89mm]{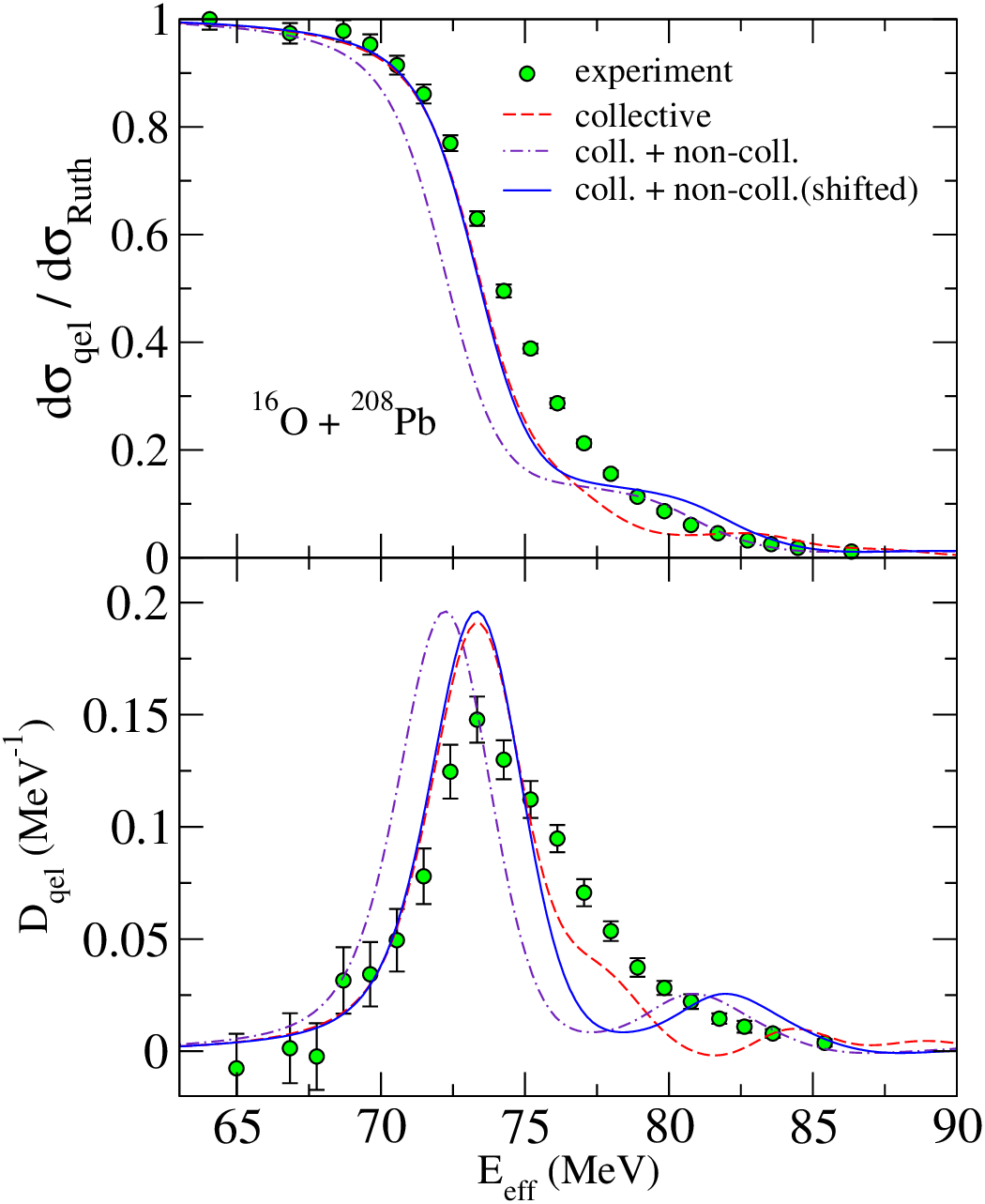}
    \caption{(Color online) Quasi-elastic scattering cross sections
(upper panel)
    and the quasi-elastic barrier distribution (lower panel) 
for the $^{16}$O +
    $^{208}$Pb system. The meaning of each line is the same as in
    Fig.~\ref{fig:fusion}. Data are taken from Ref.~\cite{T96}.}
    \label{fig:qel}
\end{figure}

Figure \ref{fig:qel} shows the quasi-elastic scattering cross section 
and the
quasi-elastic barrier distribution, $D_{\rm qel}(E)= d[\sigma_{\rm
qel}/\sigma_{\rm_R}]/dE$ at 
$\theta_{\rm cm}=170^{\circ}$.
$E_{\rm eff}$ is the effective energy defined by~\cite{timmers,HR04}
\begin{eqnarray}
E_{\rm eff} = 2E\frac{{\rm sin}(\theta/2)}{1+{\rm sin}(\theta/2)},
\end{eqnarray}
which takes into account the centrifugal energy for the Rutherford trajectory.
The meaning of each line is the same as in Fig.~\ref{fig:fusion}.
The solid lines are shifted in energy with the same amount as in the fusion
calculation.
The experimental data are taken from Ref.~\cite{T96}.

One can observe 
that the change in the barrier distribution 
due to the non-collective excitations is 
similar to the fusion calculation. 
That is, 
the main effect of the non-collective excitations is 
the barrier renormalization 
without changing the shape of 
the distribution, although they smear the barrier 
distributions at 
relatively higher energies. 
The agreement with the experimental data around 
$E_{\rm eff}=75$~MeV is not improved by the non-collective excitations. 

\subsection{Double phonon calculation}

We next show the results for the calculations with the double
octupole phonon excitations 
in $^{208}$Pb.  
In this case, the number of channels included amounts to 148.
The double octupole phonon states in $^{208}$Pb have been experimentally
investigated in Ref.~\cite{YGMY96, VMA97, YKG98, VMC98, VPE01} and 
candidates for the double phonon have been 
identified. In the present calculation we assume, for simplicity, that
all four double octupole phonon states are degenerate with $E$=5.23~MeV, 
that is, twice the energy of the single-phonon state.

\begin{figure}[t]
    \includegraphics[clip,width=71.46mm,height=156.20mm]{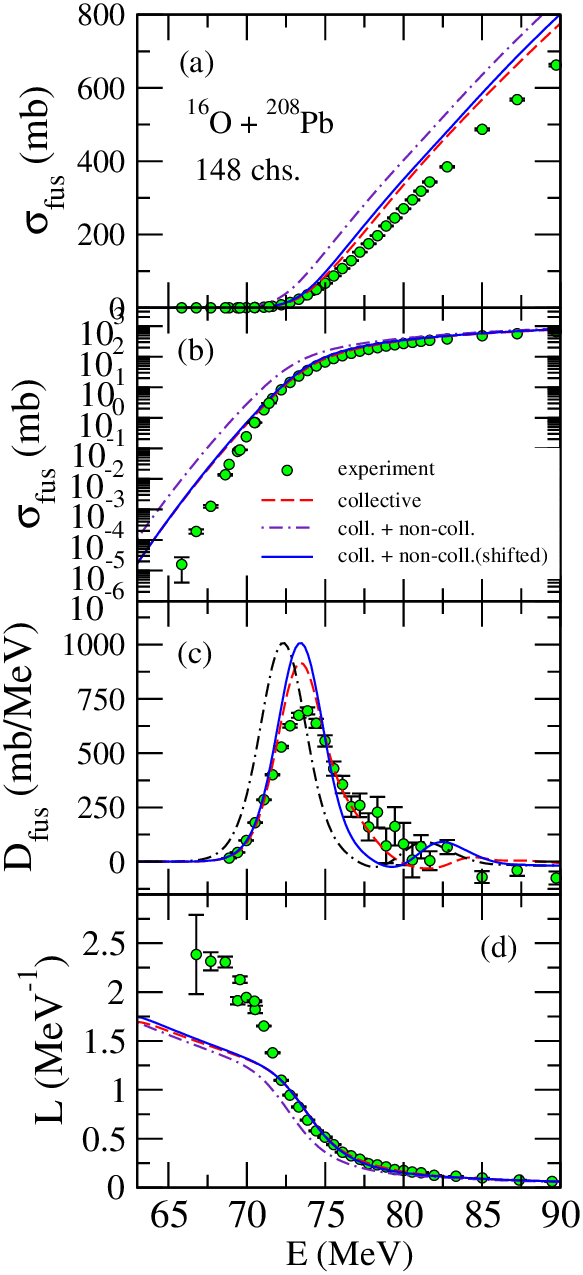}
    \caption{(Color online) Same as Fig.~\ref{fig:fusion}, 
but with the double octupole
    phonon excitations. }
    \label{fig:fusion2}
\end{figure}

\begin{figure}[t]
    \includegraphics[clip,width=85mm,height=103.89mm]{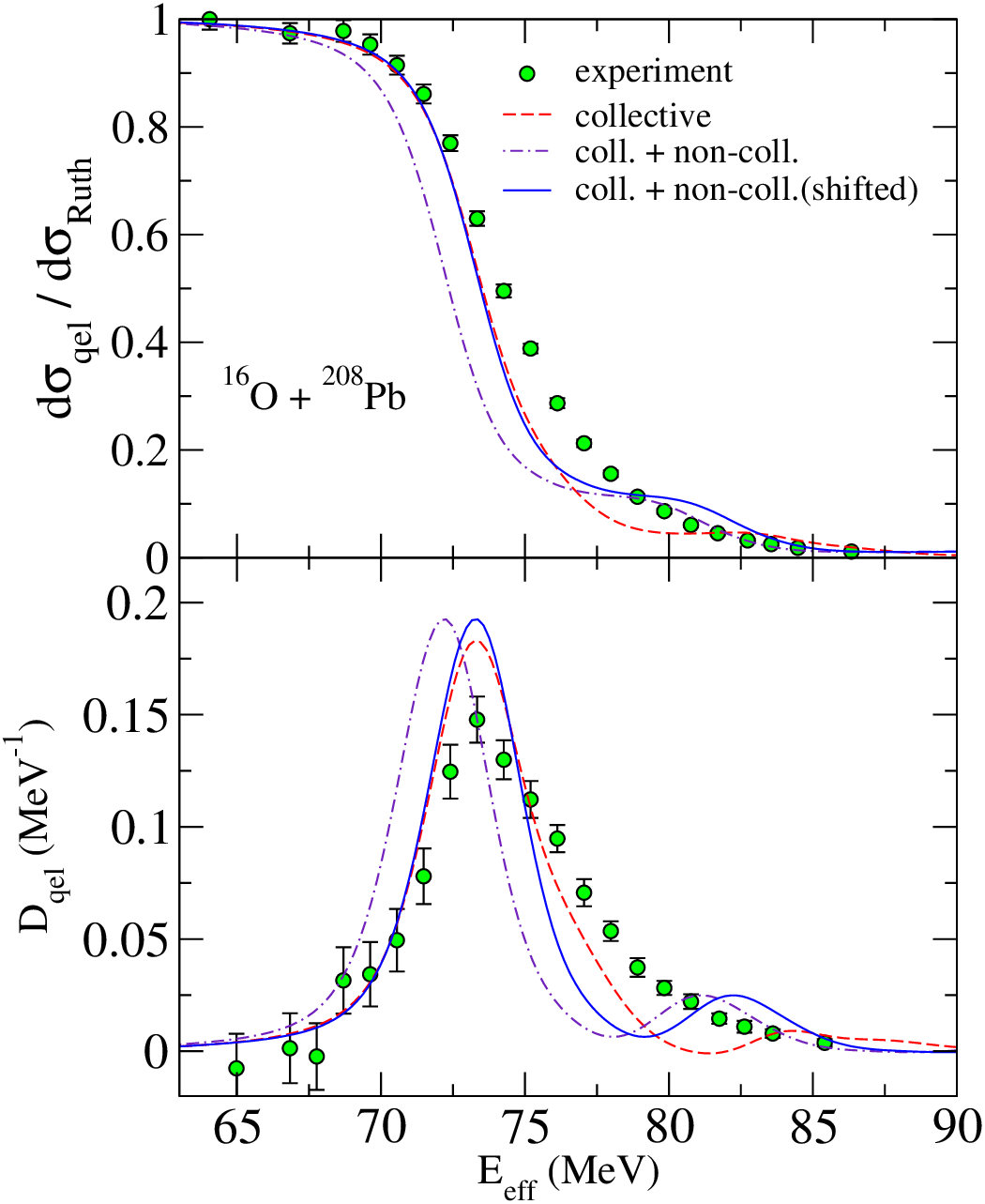}
    \caption{(Color online) Same as Fig.~\ref{fig:qel}, but with 
the double octupole
    phonon excitations.}
    \label{fig:qel2}
\end{figure}

In Figs.~\ref{fig:fusion2} and \ref{fig:qel2}, we show the calculations
for the fusion reaction and quasi-elastic scattering, respectively.
One sees that the double phonon excitations leads only to a minor 
improvement both for fusion and quasi-elastic scattering. 
The effects of the non-collective excitations are 
similar to those in the single-phonon case presented in the 
previous subsection. 
That is, the barrier distribution is smeared above the barrier while
the shape of the lower peak is almost unchanged.

We have also investigated the role of anharmonicity of the octupole 
phonon excitations of 
$^{208}$Pb~\cite{HTK97,ZH08}, together with the non-collective 
excitations. 
We have found that the effect of anharmonicity is 
minor 
and 
again does not improve the agreement with the data. 

\subsection{Q-value distribution}

\begin{figure}[t]
    \includegraphics[clip,width=86mm,height=77.26mm]{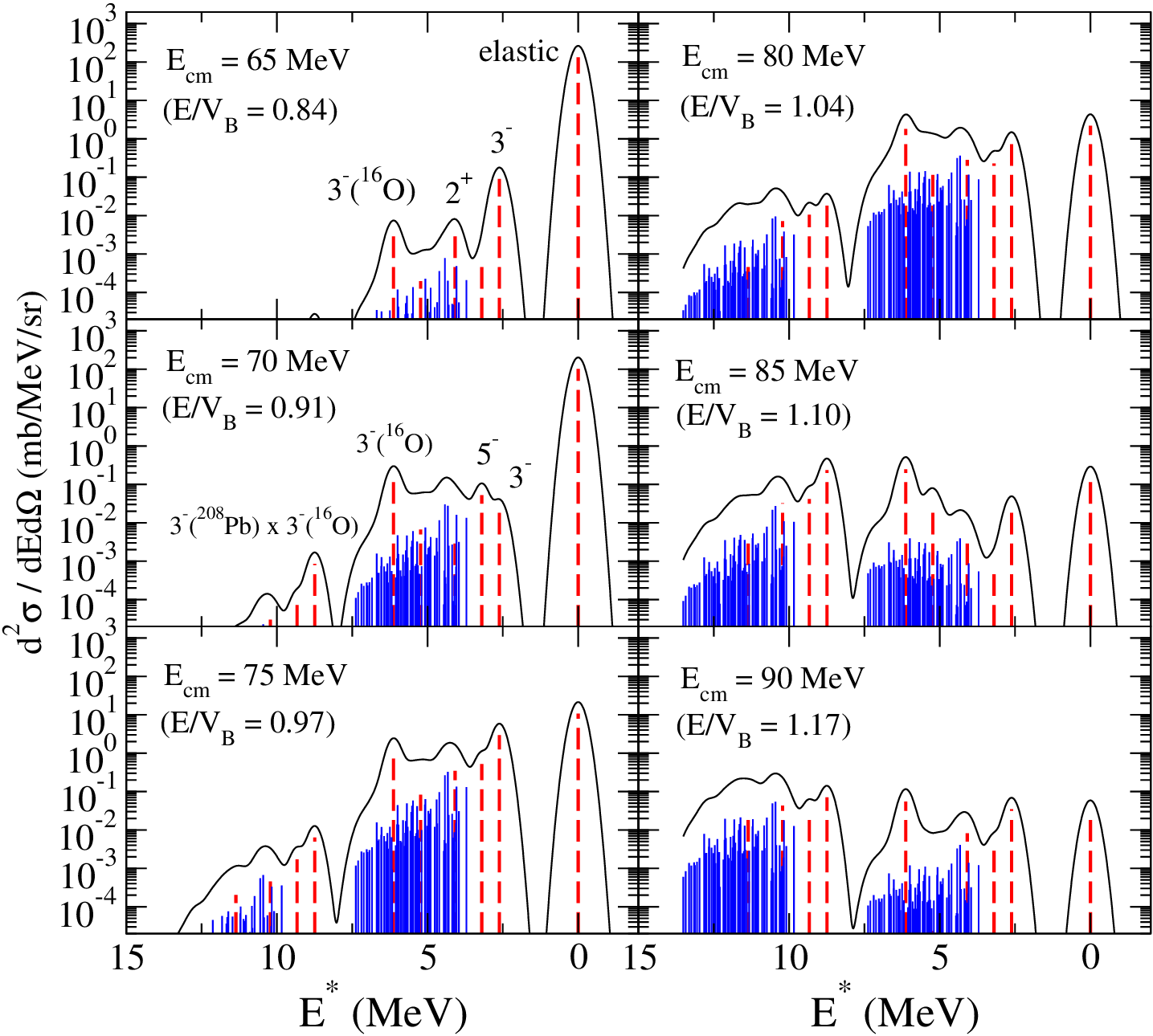}
    \caption{(Color online)The Q-value spectra for the quasi-elastic scattering
    at $\theta_{\rm cm}$=170$^{\circ}$
    for the $^{16}$O + $^{208}$Pb system for six different incident energies. The
    dashed peaks correspond to the collective excitations while the solid peaks
    correspond to the non-collective excitations. The solid line is obtained by
    smearing the peaks with a gaussian function.}
    \label{fig:qdist}
\end{figure}

Measurements of the Q-value distribution for backward-angle quasi-elastic 
scattering have been performed for this system~\cite{evers,lin}.  
The experimental data indicate that the contribution
from the non-collective excitations increases as the incident 
energy increases. 
A big advantage of our method is that the Q-value distribution can be computed 
easily because we explicitly take into account the non-collective excitations 
in our coupled-channels calculations. 

Figure \ref{fig:qdist} shows the Q-value distributions
at $\theta_{\rm cm}=170^{\circ}$ at six different incident
energies, corresponding to the double phonon calculations 
shown in Sec. III B. 
The spectra shown by the dashed lines 
correspond to the collective excitations while those by the
solid lines correspond to the non-collective excitations.
The envelope of the spectra is obtained 
by smearing with a gaussian function, 
\begin{eqnarray}
F(E^{*}) =
\sum_{n}\frac{d\sigma_n}{d\Omega}\,\frac{1}{\sqrt{2\pi}\Delta}e^{-\frac{(E^*-\epsilon_n)^2}{2\Delta^2}}, 
\end{eqnarray}
with $\Delta=0.2$~MeV. 

Note that we include the non-collective states of $^{208}$Pb up to 
7.382~MeV. Thus the spectra above this energy correspond to
mutual excitations of the $^{208}$Pb and $^{16}$O nuclei.
One can see that, at the lowest incident energy shown in the figure, the 
contribution from the collective channels is dominant.
With increasing energy, the contribution from the non-collective
excitations becomes more and more important.
This behaviour is consistent with the experimental Q-value
distribution for this system~\cite{evers,lin}.

Note that this energy dependence is 
also related to how the non-collective excitations 
modify the energy dependence of the 
barrier distribution.
Namely, at low energies where the contribution from the
non-collective excitations is not important, a change 
in the barrier distribution
is not observed. 
On the other hand, at higher energies where the
contribution from the non-collective excitations is important, the barrier
distribution is smeared due to the non-collective excitations.

\subsection{Mass-number dependence of the effect of non-collective 
excitations}

Finally, we investigate how the effect of non-collective excitations 
depends
on the mass number of the projectile nucleus. 
For this purpose, we solve the 
coupled-channels equations for the $^{32}$S + $^{208}$Pb and
$^{40}$Ca + $^{208}$Pb systems. For the nuclear potential, 
we use the Aky\"{u}z-Winther potential~\cite{Akyuz-Winther}. 
We include the same excited states in the $^{208}$Pb nucleus as those 
in the calculation for the $^{16}$O + $^{208}$Pb system discussed in the 
previous subsections. 

\begin{figure}[t]
    \includegraphics[clip,width=71.46mm,height=121.74mm]{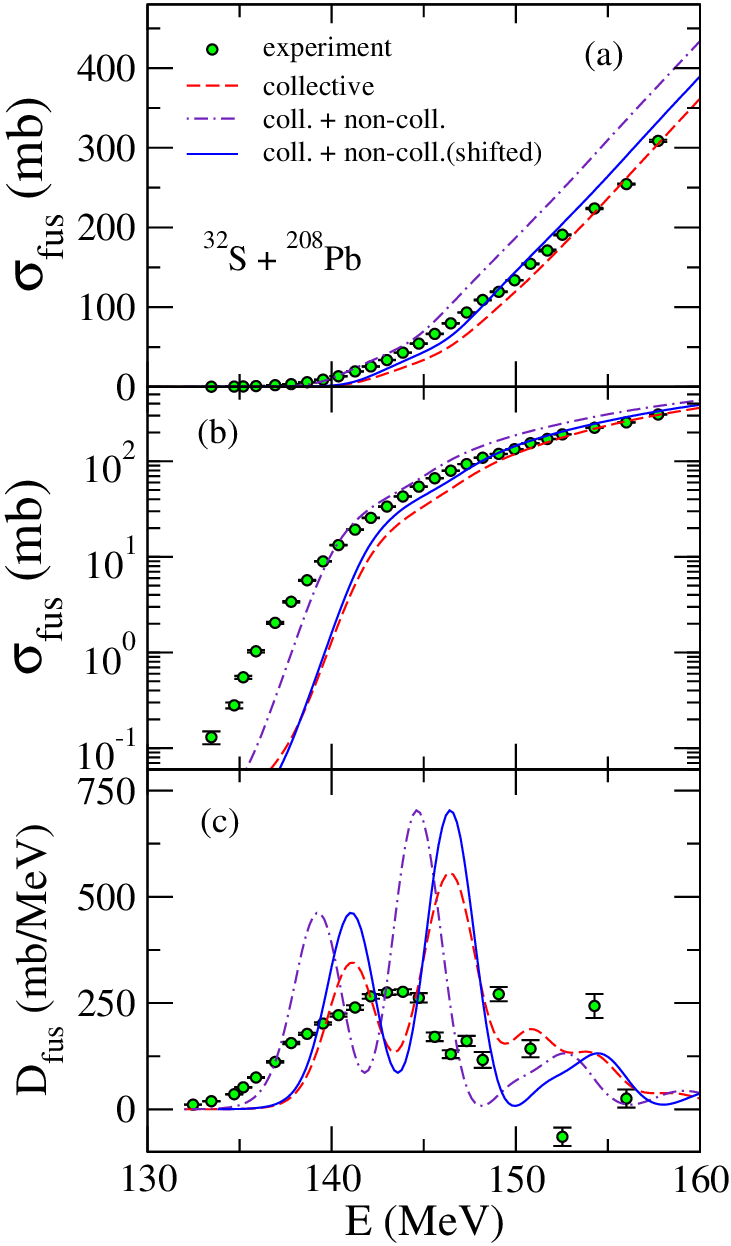}
    \caption{(Color online) Fusion cross section and fusion barrier 
distribution
    for the $^{32}$S + $^{208}$Pb system. The meaning of each line is the same as
    in Fig.~\ref{fig:fusion}. The experimental data are taken from Ref. 
~\cite{T96}.}
    \label{fig:fusion3}
\end{figure}

\begin{figure}[t]
    \includegraphics[clip,width=71.46mm,height=121.58mm]{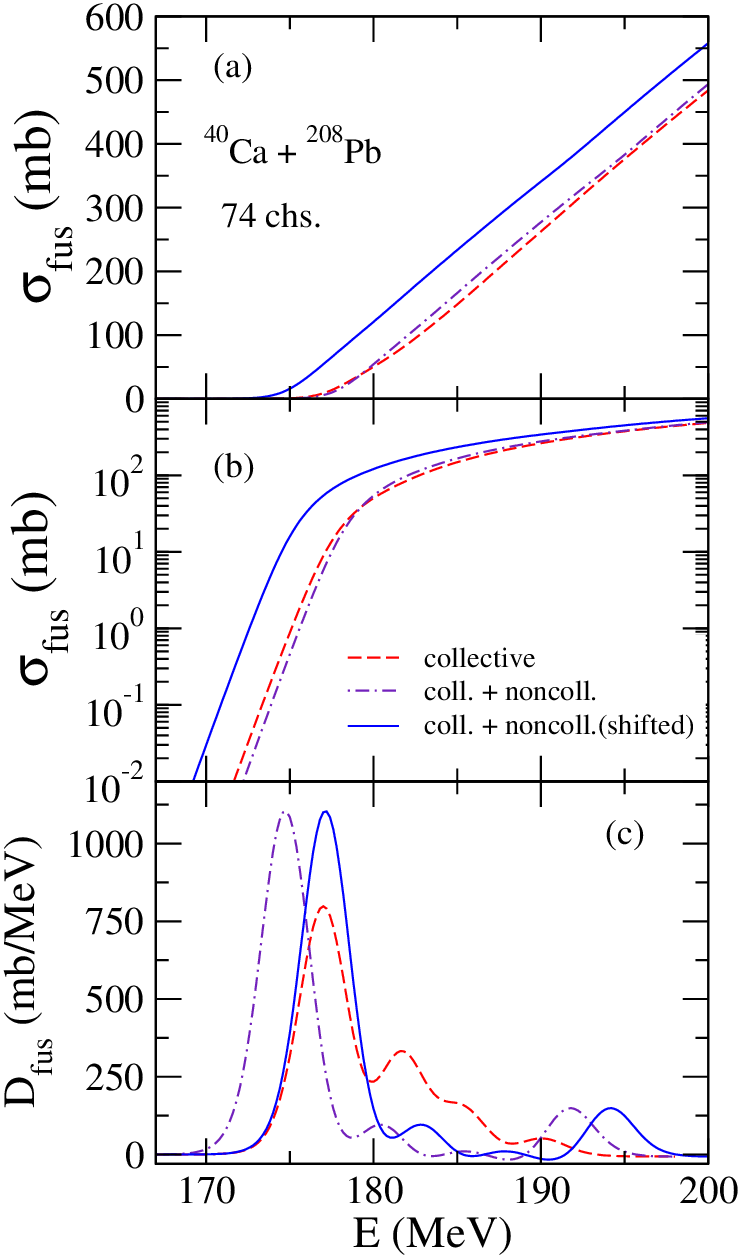}
    \caption{(Color online) Fusion cross section and fusion barrier distribution
    for the $^{40}$Ca + $^{208}$Pb system. The meaning of each line is the same as
    in Fig.~\ref{fig:fusion}.}
    \label{fig:fusion4}
\end{figure}

We first discuss the $^{32}$S + $^{208}$Pb reaction. 
For the excitations of $^{32}$S, we take into account the 
quadrupole vibration up
to double phonon states. The excitation energy and the 
deformation parameter are taken from Ref.~\cite{RNT01}.
Figure \ref{fig:fusion3} shows the calculated fusion cross section and fusion
barrier distribution. The meaning of each line is the same as in Fig.~1. 
The experimental data are taken from Ref.~\cite{T96}.
One can see that the effect of the non-collective excitations
is qualitatively similar to that in the $^{16}$O + $^{208}$Pb reaction. That is, the
barrier is shifted towards lower energy and the higher part of the barrier
distribution is smeared. However, the smearing is stronger
than that in the $^{16}$O + $^{208}$Pb system, because 
an effective coupling strength 
is in general approximately proportional to the 
charge product of the colliding nuclei~\cite{RSS91}, and thus 
the non-collective excitations are effectively stronger
for heavier systems. 
One can also see that the two low-energy 
peaks in the barrier distribution are sharpened due to the non-collective 
excitations, while the separation between the peaks is not altered much. 
The calculations do not reproduce the experimental data, 
and this might be attributed to the role of transfer reactions. 

Figure \ref{fig:fusion4} shows the fusion cross section and the fusion barrier 
distribution for the $^{40}$Ca + $^{208}$Pb reaction. 
For this system, we assume that $^{40}$Ca is inert and take into account
only the excitations of $^{208}$Pb.
As the charge product is larger, 
the effect of the non-collective
excitations is stronger than that in the $^{16}$O + $^{208}$Pb and
$^{32}$S + $^{208}$Pb reactions. It smears the higher part of the barrier
distribution while the lower main peak is sharpened.

As we have shown, while the effect of non-collective excitations is not 
large for the $^{16}$O+$^{208}$Pb system, the effect becomes increasingly 
important for heavier systems, such as $^{40}$Ca+$^{208}$Pb. 
This suggests that the conventional coupled-channels approach, that 
neglects the non-collective excitations, is well justified for 
relatively light systems, but the non-collective excitations have to 
be included explicitly in coupled-channels calculations 
for heavy-systems, for example, those relavant to a synthesis of superheavy 
elements.

\section{Summary}

We have solved the coupled-channels equations for fusion and
quasi-elastic scattering for the $^{16}$O + $^{208}$Pb system, 
including both the collective and non-collective excitations of $^{208}$Pb. 
Non-collective excitations are approximated by
vibrational couplings, whose coupling strength and excitation energy 
are taken from the analysis of the high-resolution proton inelastic 
scattering experiment.

Our results show that the barrier distributions for the fusion 
reaction and the
quasi-elastic scattering are changed in a similar manner 
due to the non-collective excitations 
at energies above the Coulomb barrier. 
The energy dependence of the cross sections, on the other hand, 
is not affected much by the non-collective excitations 
and the degree of agreement with the
experimental barrier distributions remains the same. 
In order to improve the agreement, one would therefore have 
to consider another mechanism, such as a reduction 
of the excitation energy of the 3$^-$ state in 
$^{208}$Pb as suggested in Ref. \cite{RH09}. 

The fusion calculations are also performed for the $^{32}$S + $^{208}$Pb and 
$^{40}$Ca + $^{208}$Pb systems in order to investigate the projectile 
mass-number dependence of the effect of the non-collective excitations. 
We have shown that the effect of the non-collective
excitations becomes stronger as the mass number of the 
projectile nucleus increases.

For the $^{32}$S + $^{208}$Pb system, the coupled-channels calculations 
with only the inelastic excitations of the colliding nuclei 
do not account for the experimental data. 
That is, the subbarrier fusion cross sections are significantly 
underestimated in this case and
the experimental barrier distribution is much more smeared than that
obtained by the coupled-channels calculation.
The transfer process, as well as the non-collective excitations, 
should be taken into account for this system
in order to improve the agreement with the data.

We also calculated the energy dependence of the Q-value distribution 
for the $^{16}$O + $^{208}$Pb system and found
that the contribution from the non-collective excitations 
becomes more and more
important as the incident energy increases. 
This behaviour is consistent with
the experimental Q-value distribution for the same system.

In this study, we have investigated systems involving the 
$^{208}$Pb nucleus because good information on its non-collective 
excitations is available.
However, in general, such information is not necessarily available 
for other systems. For the $^{90,92}$Zr nuclei, several proton inelastic scattering 
experiments have been performed~\cite{SK66, DES68, BSG83}. 
However, the number of identified levels 
is not as large as for $^{208}$Pb. 
Thus, we will have to use an approach different
from this work, for example the random matrix theory~\cite{YHR10}, 
to describe the non-collective excitations in the $^{20}$Ne
+ $^{90,92}$Zr system.

\begin{acknowledgments}
This work was supported by the Global COE Program
``Weaving Science Web beyond Particle-Matter Hierarchy'' at
Tohoku University,
and by the Japanese
Ministry of Education, Culture, Sports, Science and Technology
by Grant-in-Aid for Scientific Research under
the program number (C) 22540262.
\end{acknowledgments}

\end{document}